\documentclass{elsart}

\usepackage{amssymb}
\usepackage{multicol}
\usepackage{float}
\usepackage[square,sort,comma,numbers]{natbib}
\usepackage{amsmath}
\usepackage{graphics}
\usepackage[latin1]{inputenc}
\usepackage[T1]{fontenc}
\usepackage{rotate}
\usepackage{color}
\usepackage{wrapfig}
\usepackage{latexsym,epsfig,fancyheadings}
\usepackage{latexsym,epsfig}
\graphicspath{{Figures/}}
\usepackage[english,francais]{babel}

\def\og{\leavevmode\raise.3ex\hbox{$\scriptscriptstyle\langle\!\langle$~}}
\def\fg{\leavevmode\raise.3ex\hbox{~$\!\scriptscriptstyle\,\rangle\!\rangle$}}

\begin{document}

\begin{frontmatter}

\selectlanguage{english}
\title{A new analytical approach for modelling the added mass and hydrodynamic interaction of two cylinders subjected to large motions in a potential stagnant fluid}

\selectlanguage{english}
\author[Affil1]{Romain Lagrange},
\ead{romain.g.lagrange@gmail.com}
\author[Affil1]{Xavier Delaune},
\author[Affil1]{Philippe Piteau},
\author[Affil1]{Laurent Borsoi},
\author[Affil2]{Jos\'e Antunes}

\address[Affil1]{Den-Service d'\'etudes m\'ecaniques et thermiques (SEMT), CEA, Universit\'e Paris-Saclay, F-91191, Gif-sur-Yvette, France}
\address[Affil2]{Centro de Ci\^encias e Tecnologias Nucleares, Instituto Superior T\'ecnico, Universidade de Lisboa, Estrada Nacional 10, Km 139.7, 2695-066 Bobadela LRS, Portugal}


\medskip
\begin{center}
{\small Received *****; accepted after revision +++++\\
Presented by *****}
\end{center}

\begin{abstract}
A potential theory is presented for the problem of two moving cylinders, with possibly different radii, large motions, immersed in an perfect stagnant fluid. We show that the fluid force is the superposition of an added mass term, related to the time variations of the potential, and a quadratic term related to its spatial variations. We provide new simple and exact analytical expressions for the fluid added mass coefficients, in which the effect of the confinement is made explicit.
The self-added mass (resp. cross-added mass) is shown to decrease (resp. increase) with the separation distance and increase (resp. decreases) with the radius ratio. 
We then consider the case in which one cylinder translates along the line joining the centers with a constant speed. We show that the two cylinders are repelled from each other, with a force that diverges to infinity at impact. We extend our approach to the case in which one cylinder is imposed a sinusoidal vibration. We show that the force on the stationnary cylinder and the vibration displacement have opposite (resp. identical) axial (resp. transverse) directions. For large vibration amplitudes, this force is strongly altered by the nonlinear effects induced by the spatial variations of the potential. The force on the vibrating cylinder is in phase with the imposed displacement and is mainly driven by the added mass term. 
The results of this paper are of particular interest for engineers who need to grab the essential features associated to the vibration of a solid body in a still fluid.


\vskip 0.5\baselineskip


}

\end{abstract}

\end{frontmatter}

\selectlanguage{english}

\section{Introduction}

The interaction between a fluid and moving bodies is a fundamental problem which finds many applications, for example in turbomachinery \cite{Furber1979}, fish schooling \cite{Nair2007}, heat exchangers tube banks \cite{Chen1975b, Chen1977}, vibration of flexible risers \cite{LeCunff2002}, biomechanics of plants \cite{DeLangre2008}, or energy harvesting \cite{Doare2011,Singh2012,Michelin2013,Virot2016} from flapping flags \cite{Eloy2008}... We refer the reader to the book of Pa\"idoussis et. al. \cite{Paidoussis2014} for a complete bibliography of important works in this field. 
A classical approach to understand the fluid dynamics surrounding the bodies is to consider a potential flow model, in which viscous, rotational vortex and wake effects are disregarded. For large Reynolds numbers and outside the boundary layers, the potential flow theory is expected to provide a good approximation of the solution, or some of its related characteritics as the added-mass, see Chen \cite{Chen1987}.
The very first discussions of potential flows around two circular cylinders are probably due to Hicks \cite{Hicks1879}, who studied the motion of a cylindrical pendulum inside another cylinder filled with fluid. To solve this problem, Hicks considered the potential due to flow singularities distributed over the fluid domain and the cylinder surfaces. The strengths and locations of the singularities were expressed as a set of iterative equations and the potential in an integral form, suitable for numerical computation \cite{Lamb1945}. Many investigators \cite{Greenhill1882, Basset1888, Carpenter1958, Birkhoff1960, Gibert1980, Landweber1991} then tried to clarify and simplify the method of singularities by Hicks to derive a more tractable expression for the fluid potential. Still, this approximate method remains difficult to apply and as the gap between the cylinders becomes small, a high density of singularities is necessary for the solution to converge. More recently, analytical approaches based on complex analysis and conformal mapping methods \cite{Wang2004, Burton2004, Tchieu2010} have overcome this problem and general theories to determine the flow past multiple bodies have been proposed \cite{Scolan2008, Crowdy2006, Crowdy2010}. The two cylinders problem can be solved in this framework, but here we provide a more flexible method which yields new analytical and simplified expressions for the added mass coefficients. We also show that the large motions of the cylinders induce some strong spatial variations of the potential which yield nonlinear inertial effects of the fluid forces.   
\\
\\
This paper is organized as follows. Section \ref{Sec.DefinitionOfTheProblem} presents the problem of two parallel circular cylinders immersed in
an inviscid fluid and introduces the
governing dimensionless numbers. In
\S~\ref{Sec:TheoreticalModel}, we solve the potential flow problem, provide analytical expressions of the added mass terms and determine the fluid force on the cylinders.
In \S~\ref{Sec:Results} we test our predictions against published results and consider the case of a vibrating cylinder.
Finally, some conclusions are drawn in \S~\ref{Sec:Conclusion}.


\section{Definition of the problem}\label{Sec.DefinitionOfTheProblem}

\begin{figure}[H]
\begin{center}
\includegraphics[width=1\textwidth]{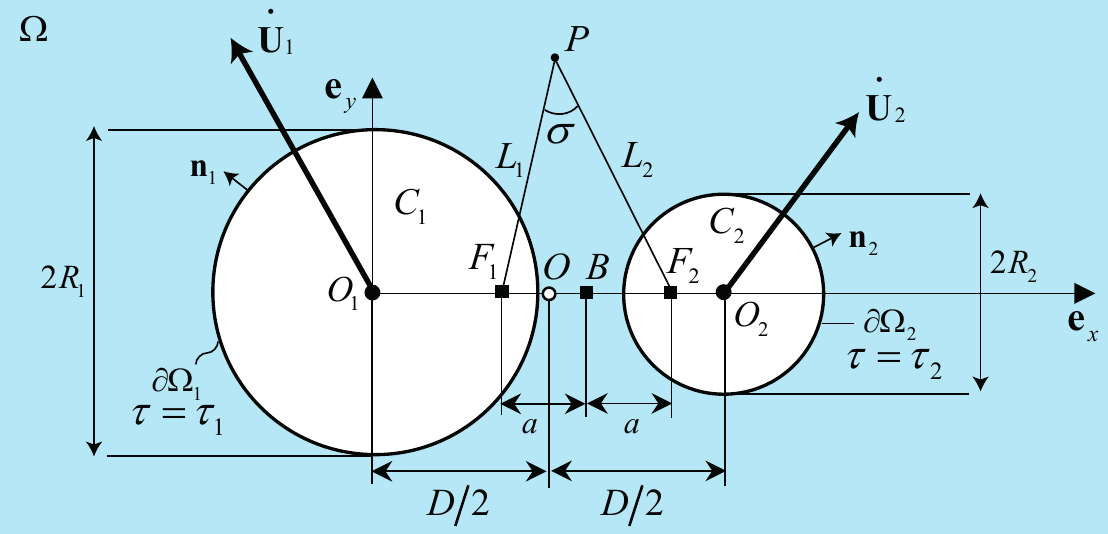}
\caption{Schematic diagram of the system: two moving cylinders ${\mathcal{C}_i}$ with radii ${R_i}$, centers ${O_i}$, velocities ${{\dot{\bf{U}}}}_i\left(T\right)$, are immersed in an inviscid fluid $\Omega$. The motions of ${\mathcal{C}_i}$ generate an incompressible and irrotational flow. The instantaneous center-to-center distance is ${D}$. The bipolar coordinates $\sigma  \in \left[ {0,2\pi } \right[$ and $\tau  = \ln \left( L_1/L_2\right)$ are used to track the position of a fluid particle $P$. The points $\left( {B,{F_1},{F_2}} \right)$ are defined in such a way that points lying on the cylinder boundaries $\partial {\Omega_i}$ have bipolar coordinates $\tau  = {\tau _i}$.}\label{Fig1}
\end{center}
\end{figure}

Let ${\mathcal{C}_i}$, $\left(i = 1,2\right)$, be two circular cylinders with radii ${R_i}$, immersed in an inviscid fluid of volume mass density $\rho $. We aim to determine the 2D flow generated by the arbitrary motions of ${\mathcal{C}_i}$, (Figure \ref{Fig1}).
For convenience we introduce rescaled quantities to reduce the number of parameters of the problem. We use ${R_2}$ to normalize lengths, some characteristic speed ${V_0}$ to normalize velocities and $\rho {V_0}^2$  to normalize pressures. We define
\begin{equation}\label{DimensionlessNumbers}
r = \frac{{{R_1}}}{{{R_2}}},{\rm{ }}{d} = \frac{{{D}}}{{{R_2}}},{\rm{ }}{\bf{u}}_i = \frac{{\bf{U}}_i}{{{R_2}}},{\rm{ }}t = \frac{{T{V_0}}}{{{R_2}}},{\rm{ }}p = \frac{P}{{\rho {V_0}^2}},{\rm{ }} {\bf{f}}_i  = \frac{{ {\bf{F}}_i }}{{\rho {V_0}^2{R_2}}},
\end{equation}
as the radius ratio, the dimensionless center-to-center distance, cylinder displacement, time, fluid pressure and fluid force, respectively. We also introduce the dimensionless separation distance, ${\varepsilon} = {d} - \left( {r + 1} \right)$ such that ${\varepsilon} = 0$ corresponds to cylinders in contact.

\section{Theoretical model}\label{Sec:TheoreticalModel}

\subsection{Fluid equations}
It is assumed that the fluid is inviscid and the flow generated by the tubes motions is incompressible and irrotational.
We thus model the flow with a potential function $\varphi $ which satisfies the Laplace equation
\begin{equation}\label{LaplaceEquation}
\Delta \varphi  = 0    \;\;\;\mbox{in $\Omega $},
\end{equation}
and the inviscid boundary conditions
\begin{equation}\label{BoundaryConditions}
\left( {\nabla \varphi  - \dot{\mathop {\bf{u}}}_i  } \right) \cdot {{\bf{n}}_i} = 0 \;\;\;\mbox{on $\partial {\Omega_i}$}.
\end{equation}
In the above eqs., $\Omega $ is the fluid domain and ${{\bf{n}}_i}$ is the outward normal unit vector to $\partial {\Omega_i}$ (Figure \ref{Fig1}). To account for an unperturbed fluid at infinity we also require that
\begin{equation}\label{EquationAtInfinity}
\varphi  \to 0   \;\;\;\mbox{at infinity.}
\end{equation}
The fluid pressure $p$ is derived from the unsteady Bernoulli equation
\begin{equation}\label{FluidPressure}
p - {p_\infty } =  - \left( {\frac{{\partial \varphi }}{{\partial t}} + \frac{{{{\left( {\nabla \varphi } \right)}^2}}}{2}} \right),
\end{equation}
where ${p_\infty }$ is the pressure at infinity. The fluid force on ${\mathcal{C}_i}$ is obtained by integration of \eqref{FluidPressure} on $\partial {\Omega_i}$
\begin{equation}\label{FluidForce}
 {\bf{f}}_i  = \int\limits_{\partial {\Omega_i}} {\left( {\frac{{\partial \varphi }}{{\partial t}} + \frac{{{{\left( {\nabla \varphi } \right)}^2}}}{2}} \right){{\bf{n}}_i}d{s_i}},
\end{equation}
where $d{s_i}$ is an infinitesimal element of integration.

The fluid equations must be solved on the instantaneous configuration to account for the possible large displacements of ${\mathcal{C}_i}$. This leads us to define a coordinate system based on the kinematics of ${\mathcal{C}_i}$ and in which the fluid eqs. can be solved analytically.

\subsection{Bipolar coordinates}
\begin{figure}[H]
\begin{center}
\includegraphics[width=0.8\textwidth]{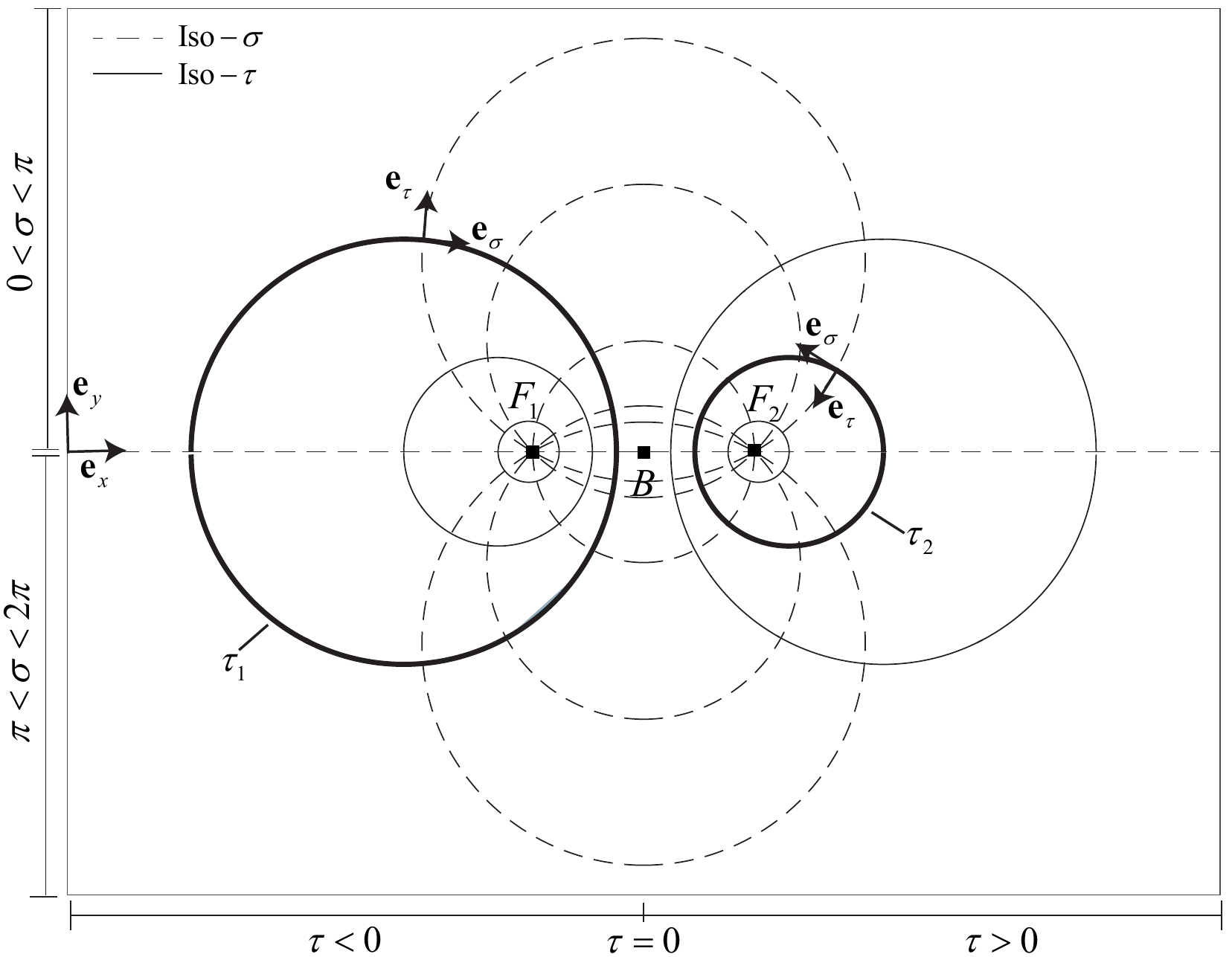}
\caption{Bipolar coordinate system. The dashed circles are iso-values of $\sigma$ while the solid circles are iso-values of $\tau$. The two foci points $F_i$ are defined such that $\partial \Omega_i$ has a coordinate $\tau_i$. The points on the $x$-axis have $\sigma=\pi$ if they lie in between $F_1$ and $F_2$, and $\sigma=0$ otherwise.}\label{FigBipolarCoor}
\end{center}
\end{figure}

Let $\left( {B,{F_1},{F_2}} \right)$ be three points defined as ${\bf{OB}} = {x_B}{{\bf{e}}_x}$ and ${\bf{B}}{{\bf{F}}_{\bf{1}}} = - a{{\bf{e}}_x}$ and ${\bf{B}}{{\bf{F}}_2} = a{{\bf{e}}_x}$, $a > 0$. From these three points, we introduce the bipolar coordinates $\sigma  \in \left[ {0,2\pi } \right[$  and $\tau  > 0$ to track the position of a fluid particle, as shown in Figure \ref{Fig1}. Since the iso-$\tau$ contours are circles with centers on the $x$-axis (Figure \ref{FigBipolarCoor}), one can align $\partial \Omega_i$ with an iso-contour $\tau_i$ by setting the position of $B$ and $F_i$. In Appendix \ref{Appendix1}, we show that 
\begin{subequations}\label{KinematicsEquationsRepere}
\begin{equation}
x_B= \left({r}^{2}-1\right)/\left(2d\right),
\end{equation}
\begin{equation}
a = {{\sqrt {{d^2} - {{\left( {1 + r} \right)}^2}} \sqrt {{d^2} - {{\left( {1 - r} \right)}^2}} } \mathord{\left/
 {\vphantom {{\sqrt {{d^2} - {{\left( {1 + r} \right)}^2}} \sqrt {{d^2} - {{\left( {1 - r} \right)}^2}} } {\left( {2d} \right)}}} \right.
 \kern-\nulldelimiterspace} {\left( {2d} \right)}},
\end{equation}
yield iso-contours for the two cylinders
\begin{equation}
{\tau _1} =  - {\sinh ^{ - 1}}\left( {{a \mathord{\left/
 {\vphantom {a r}} \right.
 \kern-\nulldelimiterspace} r}} \right) \;\;\; \mbox{and} \;\;\; {\tau _2} = {\sinh ^{ - 1}}\left( a \right).
\end{equation}
\end{subequations}
Introducing $\left( {{{\bf{e}}_\sigma },{{\bf{e}}_\tau }} \right)$ as the coordinate system basis, see Figure \ref{FigBipolarCoor} and Appendix \ref{Appendix1}, the outward normal unit vector ${{\bf{n}}_i}$  to $\partial {\Omega_i}$ is
\begin{equation}\label{NormalVectors}
{{\bf{n}}_i} = \left(-1\right)^{i-1}{{\bf{e}}_\tau }\left( {\sigma ,{\tau _i}} \right),
\end{equation}
and the infinitesimal measure $d{s_i}$ of $\partial \Omega_i$ is
\begin{equation}\label{ElementOfIntegration}
d{s_i} ={\kappa _{\sigma \tau_i }} d\sigma,
\end{equation}
with 
${\kappa _{\sigma \tau }} = {a \mathord{\left/
 {\vphantom {a {\left( {\cosh \left( \tau  \right) - \cos \left( \sigma  \right)} \right)}}} \right.
 \kern-\nulldelimiterspace} {\left( {\cosh \left( \tau  \right) - \cos \left( \sigma  \right)} \right)}}$.
 
Finally, the gradient and the Laplace operators in bipolar coordinates are
\begin{subequations}\label{DifferentialOperators}
\begin{equation}
\nabla \phi  = \frac{1}{{{\kappa _{\sigma \tau }}}}\left( {\frac{{\partial \phi }}{{\partial \sigma }}{{\bf{e}}_\sigma } + \frac{{\partial \phi }}{{\partial \tau }}{{\bf{e}}_\tau }} \right),
\end{equation}
\begin{equation}
\Delta \phi  = {\left( {\frac{1}{{{\kappa _{\sigma \tau }}}}} \right)^2}\left( {\frac{{{\partial ^2}\phi }}{{\partial {\sigma ^2}}} + \frac{{{\partial ^2}\phi }}{{\partial {\tau ^2}}}} \right).
\end{equation}
\end{subequations}

\subsection{Solution of the fluid equations}
In bipolar coordinates the fluid eqs. \eqref{LaplaceEquation}-\eqref{EquationAtInfinity} read
\begin{subequations}\label{FluidEquations}
\begin{align}
\frac{{{\partial ^2}\varphi }}{{\partial {\sigma ^2}}} + \frac{{{\partial ^2}\varphi }}{{\partial {\tau ^2}}} &= 0 \;\;\;\mbox{in $\Omega $,}\label{LaplaceEquationBipolar}\\
\frac{{\partial \varphi }}{{\partial \tau }} &= {g_{ix}}{\dot{\mathop u}_{ix}} + {g_{iy}}{\dot{\mathop u}_{iy}} \;\;\;\mbox{on $\tau  = {\tau_i}$,} \;i=1,2,\label{BoundaryFixed}\\
\varphi  &\to 0 \;\;\;\mbox{as $\left( {\sigma ,\tau } \right) \to \left( {0,0} \right)$,}\label{BoundaryInf}
\end{align}
\end{subequations}
with $\dot{u}_{ix}$, $\dot{u}_{iy}$ the ${\bf{e}}_x$ and ${\bf{e}}_y$ components of ${\dot{\bf{u}}}_i$, respectively. In \eqref{FluidEquations}, ${g_{ix}}$  and ${g_{iy}}$  are the $2\pi $  periodic functions with Fourier components  ${g_{in}} =  - 2na{e^{ - n{\left|{\tau _i}\right|}}}$ 
\begin{subequations}
\begin{align}
{g_{ix}}\left( \sigma  \right) &={\kappa _{\sigma \tau_i }} {{{{\bf{e}}_x} \cdot {{\bf{e}}_\tau }\left( {\sigma ,{\tau _i}} \right)}} =  - a\frac{{\cos \left( \sigma  \right)\cosh \left( {{\tau _i}} \right) - 1}}{{{{\left( {\cosh \left( {{\tau_i}} \right) - \cos \left( \sigma  \right)} \right)}^2}}} = \sum\limits_{n = 1}^\infty  {{g_{in}}\cos \left( {n\sigma } \right)} ,\\
{g_{iy}}\left( \sigma  \right) &={\kappa _{\sigma \tau_i }} {{{{\bf{e}}_y} \cdot {{\bf{e}}_\tau }\left( {\sigma ,{\tau _i}} \right)}}  =  - a\frac{{\sin \left( \sigma  \right)\sinh \left( {{\tau _i}} \right)}}{{{{\left( {  \cosh \left( {{\tau _i}} \right) - \cos \left( \sigma  \right)} \right)}^2}}} = \sum\limits_{n = 1}^\infty  {{g_{in}}\sin \left( {n\sigma } \right)}.
\end{align}
\end{subequations}

The fluid problem being linear in $\dot{\mathop {u}}_{ix}$ and $\dot{\mathop {u}}_{iy}$, the potential ${\varphi}$ is a superposition of two functions ${\varphi _x}$  and ${\varphi _y}$, solutions of \eqref{FluidEquations} for ${\dot{\mathop u} _{iy}} = 0$ and ${\dot{\mathop u} _{ix}} = 0$, respectively. The calculus of $\varphi_x$ and $\varphi_y$ is reported in Appendix \ref{AppendixPotential} for the sake of clarity. We find that the potential is  
\begin{align}
\varphi  &= {\dot{\mathop u}_{1x}}\sum\limits_{n = 1}^\infty  {{{\phi _{1n}}}\left[\cos \left( {n\sigma } \right)\cosh \left( {n\left( {\tau  - {\tau _2}} \right)} \right)-   \cosh \left( {n{\tau _2}} \right)\right]}\nonumber\\
&+{\dot{\mathop u}  _{1y}}\sum\limits_{n = 1}^\infty  {{{\phi _{1n}}}\sin \left( {n\sigma } \right)\cosh \left( {n\left( {\tau  - {\tau _2}} \right)} \right)}\nonumber\\
&+{\dot{\mathop u}_{2x}}\sum\limits_{n = 1}^\infty  {{{\phi _{2n}}}\left[\cos \left( {n\sigma } \right)\cosh \left( {n\left( {\tau  - {\tau _1}} \right)} \right)-   \cosh \left( {n{\tau _1}} \right)\right]}\nonumber\\
&+{\dot{\mathop u}  _{2y}}\sum\limits_{n = 1}^\infty  {{{\phi _{2n}}}\sin \left( {n\sigma } \right)\cosh \left( {n\left( {\tau  - {\tau _1}} \right)} \right)}\nonumber\\
&=\dot{\bf{u}}_1\cdot{\bf{h}}_1+\dot{\bf{u}}_2\cdot{\bf{h}}_2,\label{EqPotential}
\end{align}
with $\phi_{in}$ given by \eqref{Appendix_eq_phiIN}.

The calculation of the fluid pressure follows from the Bernoulli equation \eqref{FluidPressure}, leading to
\begin{equation}\label{PressureBipolar}
p - {p_\infty } =  - \left( \ddot{\bf{u}}_1 \cdot {\bf{h}}_1+\ddot{\bf{u}}_2 \cdot {\bf{h}}_2 + {p_q} \right),
\end{equation}
with
\begin{align}
{p_q} &= \dot{\bf{u}}_1\cdot \frac{\partial {\bf{h}}_1}{\partial t}  +\dot{\bf{u}}_2\cdot \frac{\partial {\bf{h}}_2}{\partial t} + \frac{{{{\left[ {\nabla\left (\dot{\bf{u}}_1\cdot{\bf{h}}_1+\dot{\bf{u}}_2\cdot{\bf{h}}_2\right) } \right]}^2}}}{2},\nonumber\\
&=  - \left( {{{ { \dot{\bf{u}}_1} }} \cdot \nabla {{\bf{h}}_1} + {{ { \dot{\bf{u}}_2} }} \cdot \nabla {{\bf{h}}_2}} \right) \cdot { { \dot{\bf{u}}_O}} + \frac{{{{\left[ {\nabla\left (\dot{\bf{u}}_1\cdot{\bf{h}}_1+\dot{\bf{u}}_2\cdot{\bf{h}}_2\right) } \right]}^2}}}{2},\label{Eqpq}
\end{align}
and ${{\dot{\bf{u}}_O}} = {{\left( {{\dot{\bf{u}}_1} + {\dot{\bf{u}}_2}} \right)} \mathord{\left/
 {\vphantom {{\left( {{{\bf{u}}_1} + {{\bf{u}}_2}} \right)} 2}} \right.
 \kern-\nulldelimiterspace} 2}$. 
The terms $\ddot{\bf{u}}_i \cdot {\bf{h}}_i$ come from the time variation of the fluid potential due to the motion of the cylinders. The term $p_q$ is quadratic in $(\dot {\bf{u}}_1,\dot {\bf{u}}_2)$ and comes from the spatial variations of the fluid potential. This variation is due to the motion of the cylinders (first right hand side term of \eqref{Eqpq}) and to the convective term of the Navier-Stokes equation (last RHS term).    
\\
The integration of \eqref{PressureBipolar} on $\partial {\Omega_i}$ yields the fluid force on $\mathcal{C}_i$
\begin{equation}\label{EqFluidForces}
\left( {\begin{array}{*{20}{c}}
{{{\bf{f}}_1}}\\
{{{\bf{f}}_2}}
\end{array}} \right) = -\pi\left[C_a\right]\left( {\begin{array}{*{20}{c}}
{{\ddot{\bf{u}}_1}}\\
{{\ddot{\bf{u}}_2}}
\end{array}} \right) + \left( {\begin{array}{*{20}{c}}
{{{\bf{f}}_{1q}}}\\
{{{\bf{f}}_{2q}}}
\end{array}} \right),
\end{equation}
with $\left[C_a\right]$ the added mass matrix
\begin{subequations}\label{AddedMass}\label{EqAddedMass}
\begin{align}
\left[C_a\right] &= \left( {\begin{array}{*{20}{c}}
{{m_{11}}}&0&m_{12} &0\\
0&{{m_{11}}}&0&{ - m_{12} }\\
m_{12} &0&{{m_{11}}}&0\\
0&{ - m_{12} }&0&{{m_{11}}}
\end{array}} \right),\\
m_{11}  &= \sum\limits_{n = 1}^\infty  {\frac{{4 n{a^2}{e^{ - 2n{\tau _2}}}}}{{\tanh \left[ {n\left( {{\tau _2} - {\tau _1}} \right)} \right]}}},\\
m_{12}   &= \sum\limits_{n = 1}^\infty  {\frac{{ - 4 n{a^2}{e^{ - n\left( {{\tau _2} - {\tau _1}} \right)}}}}{{\sinh \left[ {n\left( {{\tau _2} - {\tau _1}} \right)} \right]}}}.
\end{align}
\end{subequations}
These are new analytical expressions for the self and cross-added mass terms $m_{11}$ and $m_{12}$.
The self-added mass $m_{11}$ relates the fluid force on $\mathcal{C}_i$ to its own acceleration. The cross-added mass $m_{12}$ relates the fluid force on $\mathcal{C}_i$ to the acceleration of $\mathcal{C}_j$, $j\neq i$. These new expressions have the advantage to make explicit the effect of the confinement on the added mass terms: $m_{11}$ and $m_{12}$ depend on $a$ and $\tau_i$, which are functions of the separation distance, according to \eqref{KinematicsEquationsRepere}. 

The terms ${{{\bf{f}}_{iq}}}$ are quadratic in $(\dot {\bf{u}}_1,\dot {\bf{u}}_2)$ and come from the integration of $p_q$ on $\partial\Omega_i$ 
\begin{equation}\label{QuadraticForce}
{ {\bf{f}}_{iq}} =   \int\limits_0^{2\pi} {{p_q}\left( {\sigma ,{\tau _i}} \right){{\bf{n}}_i }{\kappa _{\sigma \tau_i }} d\sigma}.
\end{equation}
These quadratic terms (computed numerically) represent the part of the fluid force due to the spatial variations of the fluid potential. 

Finally, note that if the cylinders are subject to an incident unsteady flow ${\bf{v}}_\infty(t)$, the fluid potential modifies to 
\begin{equation}
\varphi  = {{\bf{v}}_\infty } \cdot {\bf{OM}} + \left( {{{\mathop {\dot{\bf{u}}} }_1} - {{\bf{v}}_\infty }} \right) \cdot {{\bf{h}}_1} + \left( {{{\mathop {\dot{\bf{u}}} }_2} - {{\bf{v}}_\infty }} \right) \cdot {{\bf{h}}_2},
\end{equation}
where ${\bf{OM}}$ is the position vector of the fluid particle $P$. The first right hand side term ensures that $\nabla \varphi  \to {\bf{v}}_\infty$ far from the cylinders. The other terms follow from \eqref{EqPotential} where ${{\mathop {\dot{\bf{u}}} }_i}-{\bf{v}}_\infty$ is the velocity of ${\mathcal{C}_i}$ in the reference frame attached to the incident flow. Noting $[I]$ the identity matrix, it comes that the fluid force is
\begin{equation}\label{Morison}
\left( {\begin{array}{*{20}{c}}
{{{\bf{f}}_1}}\\
{{{\bf{f}}_2}}
\end{array}} \right) = \pi \left( {\left[ {{C_a}} \right] + \left[ I \right]} \right)\left( {\begin{array}{*{20}{c}}
{{{\dot{\bf{v}}}_\infty }}\\
{{{\dot{\bf{v}}}_\infty }}
\end{array}} \right) - \pi \left[ {{C_a}} \right]\left( {\begin{array}{*{20}{c}}
{{{\ddot{\bf{u}}}_1}}\\
{{{\ddot{\bf{u}}}_2}}
\end{array}} \right) + \left( {\begin{array}{*{20}{c}}
{{{\bf{f}}_{1q}}}\\
{{{\bf{f}}_{2q}}}
\end{array}} \right),
\end{equation}
where $p_q$ and ${\bf{f}}_{iq}$ are given by \eqref{Eqpq}-\eqref{QuadraticForce} under the change ${{\mathop {\dot{\bf{u}}} }_i}\to {{\mathop {\dot{\bf{u}}} }_i}-{\bf{v}}_\infty$.
The equation \eqref{Morison} is similar to the Morison equation \cite{Morison1950}, which gives the inline force on a body subject to an unsteady flow. From an experimental stand point, \eqref{Morison} shows that $[C_a]$ can be extracted either from the force on tubes vibrating in a fluid at rest or from the force on fixed tubes subject to an unsteady fluid flow.  

\section{Results}\label{Sec:Results}
\subsection{Added mass}\label{Subsection_Added_Mass} 
\begin{figure}[H]
\begin{center}
\includegraphics[width=1\textwidth]{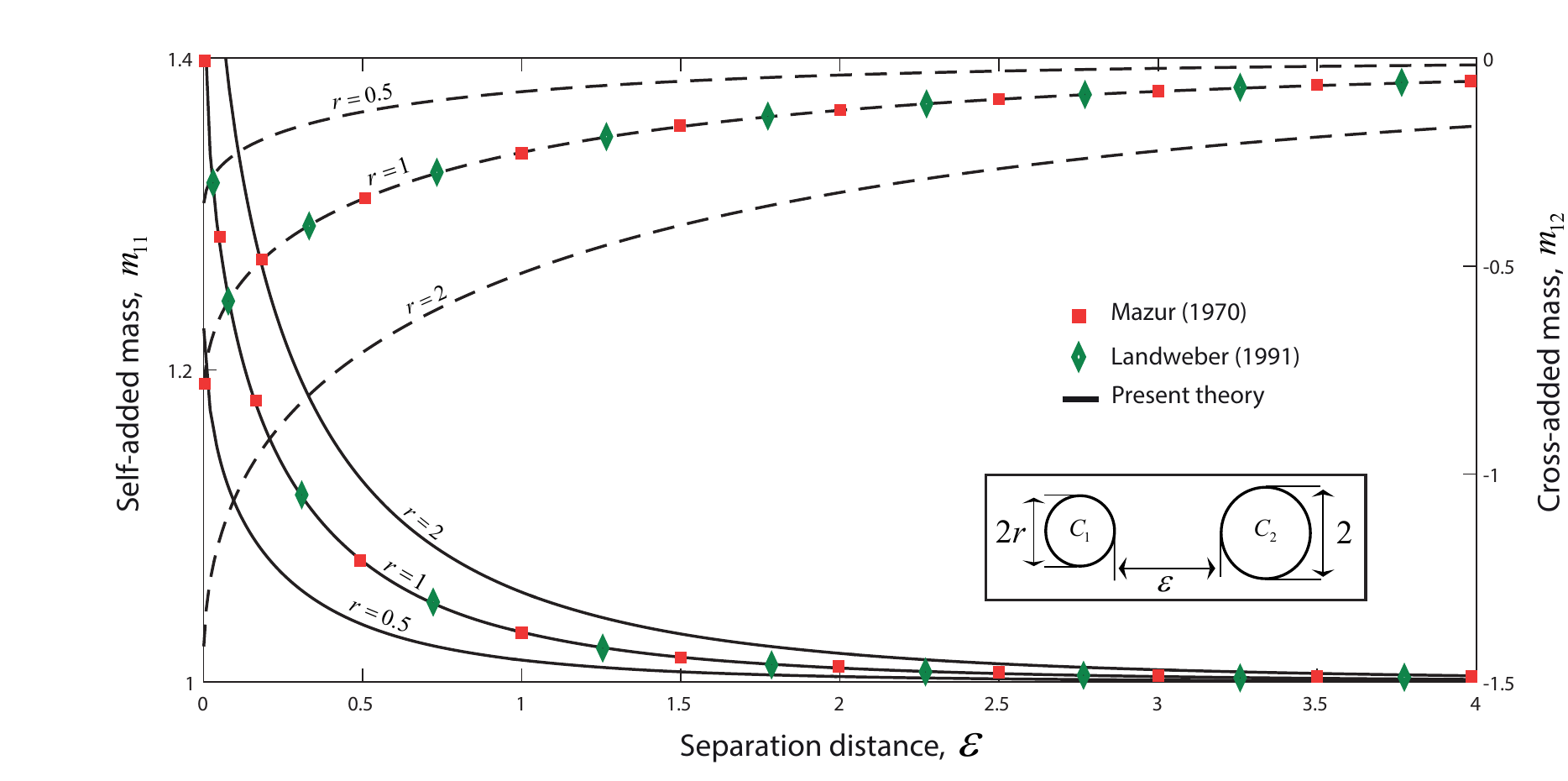}
\caption{Self-added mass ${m_{11}}$ (solid lines) and cross-added mass ${m_{12}}$, given by eq. \eqref{AddedMass}, as functions of the separation distance $\varepsilon $ and the radius ratio $r$.}\label{Fig2}
\end{center}
\end{figure}
\begin{figure}[H]
\begin{center}
\end{center}
\end{figure}
In the previous section, we have obtained new simple analytical expressions for the added mass coefficients $m_{11}$ and $m_{12}$. Their evolution with the separation distance $\varepsilon$ and the radius ratio $r$ is shown in Figure \ref{Fig2}. 
We find that $m_{11} $ (resp. $m_{12}$) decreases (resp. increases) monotonically with the dimensionless separation distance, and increases (resp. decreases) with the radius ratio. When the cylinders are in close proximity, i.e. ${\varepsilon} \to 0$, the confinement is maximum and the added mass coefficients become unbounded, as expected. When the two cylinders are far apart, i.e. ${\varepsilon} \to \infty $, they both behave like an isolated cylinder in an infinite fluid domain, $m_{11}  \to 1 $ and $m_{12} \to 0 $. To validate our observations, we have reported in Figure \ref{Fig2} the predictions of the literature \cite{Mazur1970,Landweber1991}, for $r = 1$. Unlike the current method, \cite{Mazur1970} used a conformal mapping method to solve the potential problem and extracted the added mass coefficients from the kinetic energy of the fluid. On his side, \cite{Landweber1991} extended the method of images by \cite{Hicks1879,Herman1887} and extracted the added mass coefficients from the fluid force acting on the cylinders. We obtain an excellent agreement with those authors (and some others reproduced in Table \ref{tab1}), 
thereby validating our prediction for $m_{11}$ and $m_{12}$.
\begin{table}[h]
   \centering 
\begin{tabular}{ccccccc}
  \hline
   & Present & Pa\"idoussis & Chen  & Suss  & Dalton et al.  & Gibert \\
    & reference & \cite{Paidoussis1984} &\cite{Chen1978} &\cite{Suss1977} &\cite{Dalton1971} &\cite{Gibert1980}\\ 
  \hline
  $m_{11}$ & 1.0319 & 1.0320 & 1.0325 & 1.0328 & 1.0313 & 1.0250\\
  $m_{12}$ & -0.2269 & -0.2269 & -0.2265 & -0.2266 & -0.2245 & -0.2250 \\
  \hline
\end{tabular}
\centering
\caption{Comparison of the added mass coefficients by the present analysis with those obtained by various other authors. The prediction by Gibert \cite{Gibert1980} is a bit off because it relies on an asymptotic expansion that is valid for large $\varepsilon$. The table is given for $\varepsilon=1$ and $r=1$.}\label{tab1}
\end{table}

\subsection{Quatratic fluid force}

\subsubsection{Uniform linear motion}
\begin{figure}[H]
\begin{center}
\includegraphics[width=1\textwidth]{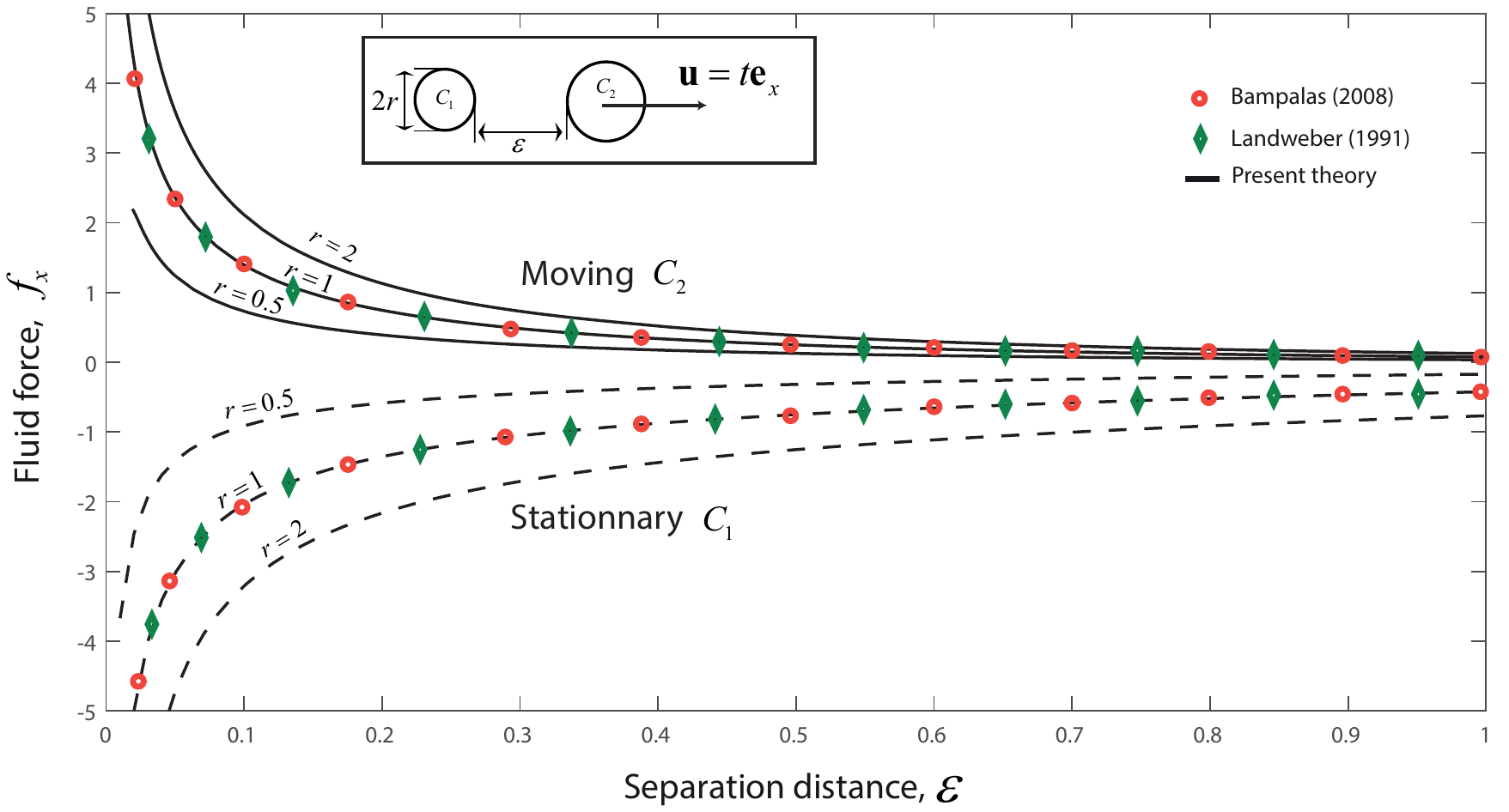}
\caption{Fluid force, computed from eq. \eqref{EqFluidForces}, as a function of the separation distance $\varepsilon $ and the radius ratio $r$. The cylinder ${\mathcal{C}_1}$ is stationnary. The cylinder ${\mathcal{C}_2}$ translates along the line joining the centers with a constant speed.}\label{Fig3}
\end{center}
\end{figure}

The fluid force is the superposition of an added mass term and a quadratic term  ${ {\bf{f}} _{iq}}$. In the previous section, we have validated the prediction for the added mass term and we now proceed with assesing the validity of ${{\bf{f}} _{iq}}$. To do so we consider the case in which ${\mathcal{C}_1}$ is stationnary while ${\mathcal{C}_2}$ translates along the line joining the centers with a constant characteristic speed ${V_0}$. The dimensionless displacement writes ${\bf{u}} = t{{\bf{e}}_{{x}}}$ with $t = T V_0/R_2$  the dimensionless time. The $y$-component of the fluid force is ${ f _{{iy}}}=0$ by symmetry.\\
\\
In Figure \ref{Fig3}, we plot ${f _{{ix}}}$, given by eq. \eqref{EqFluidForces}, as a function of the separation distance  $\varepsilon $.
We find that ${f _{{1x}}}$ (resp. ${f _{{2x}}}$) is negative (resp. positive) so that the cylinders are repelled from each other. The magnitude of this repelling force decreases monotonically with $\varepsilon $, and increases with the radius ratio $r$. When the cylinders are near to contact, the force ${f _{{ix}}}$ becomes singular and diverges to infinity. When the cylinders are far apart, the d'Alembert's paradox is at play so that no fluid force is exerted on $\mathcal{C}_i$.
In Figure \ref{Fig3} the results by \cite{Bampalas2008} and \cite{Landweber1991} are presented for comparison. In \cite{Bampalas2008} the force is computed using a M\"obius conformal mapping and a series of image singularities. Here again, we find an excellent agreement with those authors 
thereby validating our expression \eqref{QuadraticForce} for ${\bf{f}}_{iq}$.

\subsubsection{Sinusoidal vibrations}
\begin{figure}[H]
\begin{center}
\includegraphics[width=1\textwidth]{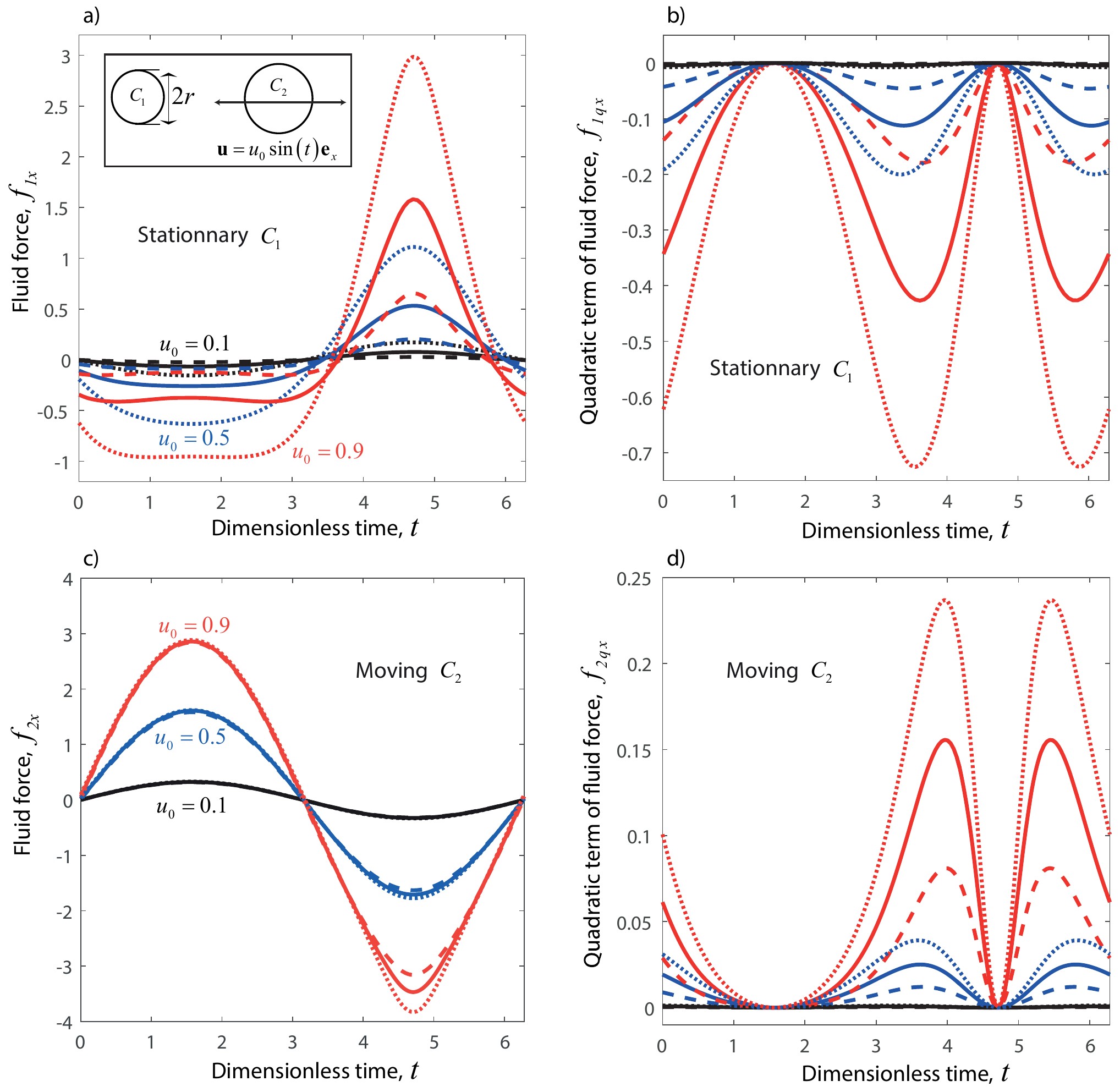}
\caption{Fluid force, computed from eq. \eqref{EqFluidForces}, as a function of the dimensionless time $t$. The cylinder ${\mathcal{C}_1}$ is stationnary. The cylinder ${\mathcal{C}_2}$ vibrates along the line joining the centers with a dimensionless amplitude ${u_0}$. The radius ratios are: $r = 0.5$ (dashed lines), $r = 1$ (solid lines), $r = 2$ (dotted lines). The dimensionless vibration amplitudes are $u_0=0.1$ (black color), $u_0=0.5$ (blue color) and $u_0=0.9$ (red color). The initial separation distance is ${\varepsilon _0} = 1$.
The temporal and spatial evolutions of the fluid pressure and velocity fields for $r=0.5$ and $u_0=0.9$ can be observed in the corresponding movies.}\label{Fig4}
\end{center}
\end{figure}
\begin{figure}[H]
\begin{center}
\includegraphics[width=1\textwidth]{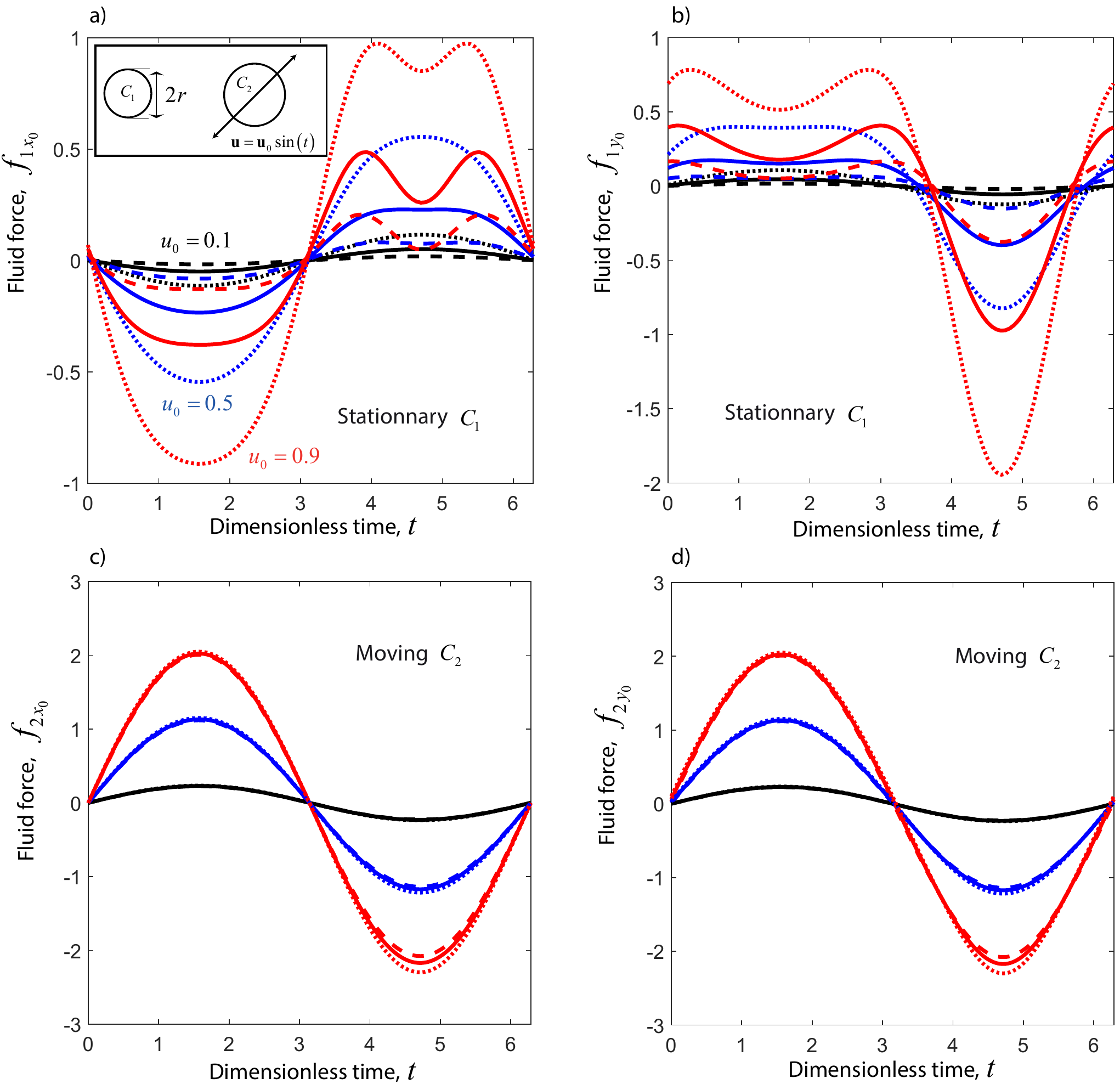}
\caption{
Fluid force, computed from eq. \eqref{EqFluidForces}, as a function of the dimensionless time $t$. The cylinder ${\mathcal{C}_1}$ is stationnary. The cylinder ${\mathcal{C}_2}$ vibrates along the direction $\left( {{{\bf{e}}_{{x_0}}},{\bf{u}}}_0 \right) = {\pi/4}$ with a dimensionless amplitude ${u_0}$. Same legend as Figure \ref{Fig4}.}\label{Appendix_Fig1}
\end{center}
\end{figure}

Thus far, we have successfully compared the results of our model to those found in the literature. We now proceed with extending those results to the case in which ${\mathcal{C}_1}$ is stationnary while ${\mathcal{C}_2}$ is imposed a sinusoidal vibration with amplitude ${U_0}$ and angular frequency $\Omega $. We choose  ${V_0} = \Omega {R_2}$ as the characteristic speed so that the dimensionless displacement writes ${\bf{u}}\left( t \right) = {u_0}\sin \left( t \right){{\bf{u}}_{{0}}}$, with ${{\bf{u}}_0} = \cos \left( \gamma  \right){{\bf{e}}_{x_0}} + \cos \left( \gamma  \right){{\bf{e}}_{y_0}}$. Here $\gamma  = \left( {{{\bf{e}}_{x_0}},{{\bf{u}}_0}} \right)$ is the angle made by the line joining the centers in the initial configuration and the direction of vibration. The initial separation distance is ${\varepsilon _0} = 1$ and three vibration amplitudes are considered, ${u_0} = \left\{ {0.1,0.5,0.9} \right\}$. The time evolution of the fluid force is shown in Figures \ref{Fig4} and \ref{Appendix_Fig1} for $\gamma=0$ and $\gamma=\pi/4$, respectively. Note that for $\gamma=0$ the problem is symmetric about the $x$-axis, such that ${f _{{iy}}}=0$ and we only show the evolution of ${f_{{ix}}}$. For $\gamma=\pi/4$ the fluid force is represented through its components $\left({ f _{{i{x_0}}}},{ f _{{i{y_0}}}}\right)$ in the initial basis $\left( {{{\bf{e}}_{x_0}},{{\bf{e}}_{y_0}}} \right)$.

For both cases ($\gamma=0$, $\gamma=\pi/4$) we find that $f_{1x}$ and $u_{2x}$ have oppposite signs, see Figures \ref{Fig4}a) and \ref{Appendix_Fig1}a). Thus, the fluid always pushes the stationnary cylinder in an axial direction that is in phase opposition with the vibration displacement. On the contrary, ${{f}}_{1y}$ and $u_{2y}$ have the same sign, see Figure \ref{Appendix_Fig1}b), such that the fluid pushes the stationnary cylinder in a transverse direction that is in phase with the imposed displacement. We also observe that ${\bf{f}}_1$ has a high harmonic distorsion due to the non-negligible effect of the quadratic term ${{\bf{f}} _{{1q}}}$, see Figure \ref{Fig4}b). It follows that the force on the stationnary cylinder is due both to the time and spatial variations of the fluid potential.      
\\
The force ${\bf{f}}_2$ on the moving cylinder has the same direction as the imposed displacement, see Figures \ref{Fig4}c) and \ref{Appendix_Fig1}c-d).   
The quadratic term ${{\bf{f}} _{{2q}}}$ remains negligible, see Figure \ref{Fig4}d), such that ${\bf{f}}_2$ has a low harmonic distorsion and approximates to ${{\bf{f}}_{2}} =-\pi m_{11} {\ddot{{\bf{u}}}_{2}}$. Thus, the force on the moving cylinder is mainly due to the time variations of the fluid potential. Finally, we note that the magnitude $|{\bf{f}}_i|$ is maximum as the distance between the cylinders is minimum, and it increases with the radius ratio. This is of course related to a confinement effect which induces an increase of the added mass terms $|m_{11}|$ and $|m_{12}|$, as already pointed out in \S~\ref{Subsection_Added_Mass}. 
\\
The online movies ($r=0.5$, $u0=0.9$) show the fluid pressure distribution and its time evolution as the cylinder vibrates. As expected, a positive (resp. negative) pressure appears in the wake of the moving cylinder as it decelerates (resp. accelerates).

\section{Conclusion}\label{Sec:Conclusion}
We have considered the problem of two moving circular cylinders immersed in an inviscid fluid. 
A potential theory based on a bipolar coordinate system shows that the fluid force on the cylinders is the superposition of an added mass term and a quadratic term. The added mass term comes from the time variations of the fluid potential while the quadratic term is related to its spatial variations. We provide new simple and exact analytical expressions for the fluid added mass coefficients, in which the effect of the confinement is made explicit. The self-added mass $m_{11}$ (resp. cross-added mass $m_{12}$) decreases (resp. increases) with the separation distance and increases (resp. decreases) with the radius ratio. When the separation distance tends to zero, the confinement is maximum and the added mass coefficients $m_{11}$ and $m_{12}$ diverge to infinity. When the two cylinders are far apart, the added mass coefficients are those of a single cylinder in a fluid of infinite extent, $m_{11}  \to 1 $ and $m_{12} \to 0$. 
We then consider the case in which one cylinder is stationnary while the other is translating along the line of centers with a constant speed. Our results indicate that the fluid forces are repulsive, diverge to infinity when the cylinders are in contact, and decrease to zero as they separate. These observations are in excellent agreement with published results. Finally, we consider the case in which one cylinder is fixed while the other is imposed a sinusoidal vibration. We show that
the force on the stationnary cylinder and the vibration displacement have opposite axial directions but identical transverse directions. This force is strongly altered by the nonlinear effects due to the spatial variations of the fluid potential. On the other hand, the force on the vibrating cylinder is mainly due to the time variations of the potential and approximates to $ {\bf{f}_2}  =  -\pi m_{11} \ddot{\mathop {\bf{u}}}_2 $.
 
\section*{Acknowledgment}
The authors would like to thank Prof. C. Eloy for his valuable comments and suggestions to improve the manuscript.

\appendix

\section{Relations between coordinates}\label{Appendix1}
\begin{figure}[H]
\begin{center}
\includegraphics[width=0.8\textwidth]{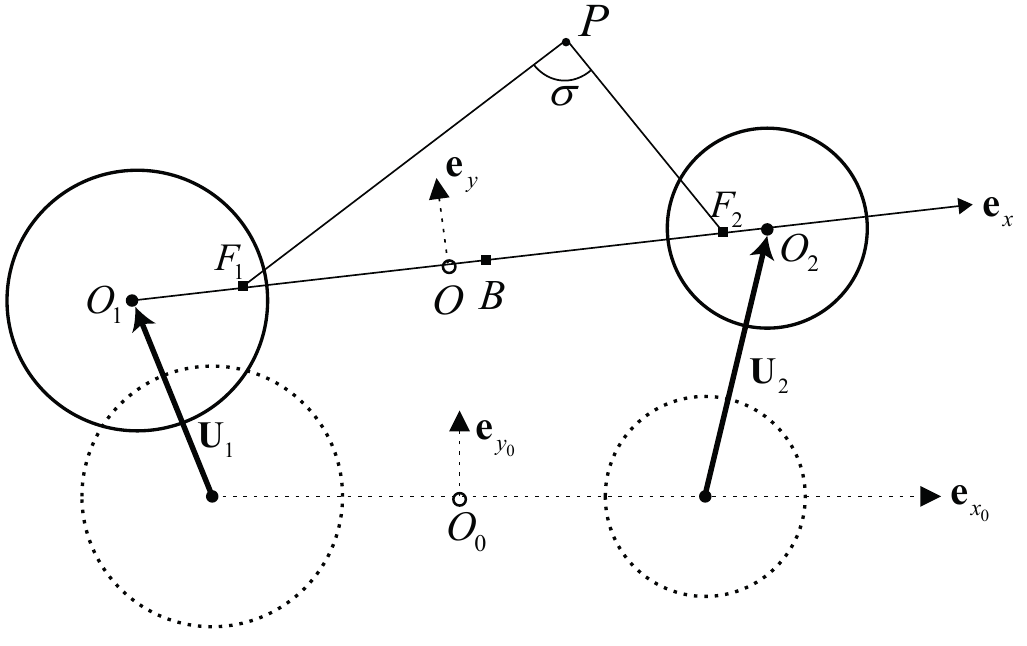}
\caption{Coordinate systems used for the two cylinders problem. The dashed lines represent the initial configuration whose frame of reference is $(O_o,{\bf{e}}_{x_0},{\bf{e}}_{y_0})$. In this frame a fluid particle $P$ has cartesian coordinates $(x_0,y_0)$. The solid lines represent the instantaneous configuration whose frame of reference is $(O,{\bf{e}}_{x},{\bf{e}}_{y})$. In this frame a fluid particle $P$ has cartesian coordinates $(x,y)$ and bipolar coordinates $(\sigma,\tau)$.}\label{AppendixCoord}
\end{center}
\end{figure}

The relations between the cartesian coordinates $\left( {{x_0},{y_0}} \right)$ (initial configuration), $\left( {x,y} \right)$ (instantaneous configuration)
and the bipolar coordinates  $\left( {\sigma ,\tau } \right)$ are, see Figure \ref{AppendixCoord}
\begin{equation}
\left( {\begin{array}{*{20}{c}}
{{x_0}}\\
{{y_0}}
\end{array}} \right) = \frac{1}{2}\left( {\begin{array}{*{20}{c}}
{{u_{{1x_0}}+u_{{2x_0}}}}\\
{{u_{{1y_0}}+u_{{2y_0}}}}
\end{array}} \right) + {\bf{P}}\left( {\begin{array}{*{20}{c}}
x\\
y
\end{array}} \right),
\end{equation}
\begin{equation}\label{eq_xy_sigmatau}
\left( {\begin{array}{*{20}{c}}
x\\
y
\end{array}} \right) = \left( {\begin{array}{*{20}{c}}
{{x_B} + {{a\sinh \left( \tau  \right)} \mathord{\left/
 {\vphantom {{a\sinh \left( \tau  \right)} {\left( {\cosh \left( \tau  \right) - \cos \left( \sigma  \right)} \right)}}} \right.
 \kern-\nulldelimiterspace} {\left( {\cosh \left( \tau  \right) - \cos \left( \sigma  \right)} \right)}}}\\
{{{a\sin \left( \sigma  \right)} \mathord{\left/
 {\vphantom {{a\sin \left( \sigma  \right)} {\left( {\cosh \left( \tau  \right) - \cos \left( \sigma  \right)} \right)}}} \right.
 \kern-\nulldelimiterspace} {\left( {\cosh \left( \tau  \right) - \cos \left( \sigma  \right)} \right)}}}
\end{array}} \right) = {\bf{g}}\left( {\sigma ,\tau } \right),
\end{equation}
\begin{equation}
\left( {\begin{array}{*{20}{c}}
\sigma \\
\tau
\end{array}} \right) = \left( {\begin{array}{*{20}{c}}
{\pi  - 2{{\tan }^{ - 1}}\left( {\frac{{2ay}}{{{a^2} - {{(x - {x_B})}^2} - {y^2} + \sqrt {{{\left( {{a^2} - {{(x - {x_B})}^2} - {y^2}} \right)}^2} + 4{a^2}{y^2}} }}} \right)}\\
{\frac{1}{2}\ln \left( {\frac{{{{\left( {x - {x_B} + a} \right)}^2} + {y^2}}}{{{{\left( {x - {x_B} - a} \right)}^2} + {y^2}}}} \right)}
\end{array}} \right),
\end{equation}
with
\begin{equation}
{\bf{P}} = \frac{1}{\sqrt {{{\left( {{d_0} + {u_{{2x_0}}-u_{{1x_0}}}} \right)}^2} + ({u_{{2y_0}}-u_{{1y_0}})}^2}}\left( {\begin{array}{*{20}{c}}
{d_0 + u_{{2x_0}}-u_{{1x_0}}}&{ - (u_{{2y_0}}-u_{{1y_0}})}\\
{u_{{2y_0}}-u_{{1y_0}}}&{d_0 + u_{{2x_0}}-u_{{1x_0}}}
\end{array}} \right),
\end{equation}
the passage matrix from the basis  $\left( {{{\bf{e}}_x},{{\bf{e}}_y}} \right)$ to $\left( {{{\bf{e}}_{{x_0}}},{{\bf{e}}_{{y_0}}}} \right)$.

From \eqref{eq_xy_sigmatau} it follows that
\begin{equation}
{y^2} + {\left[ {x - \left( {{x_B} + a\coth \left( \tau  \right)} \right)} \right]^2} = {\left( {{a \mathord{\left/
 {\vphantom {a {\sinh \left( \tau  \right)}}} \right.
 \kern-\nulldelimiterspace} {\sinh \left( \tau  \right)}}} \right)^2},
\end{equation}
so that iso-$\tau$ contours are circles of radius ${{a \mathord{\left/
 {\vphantom {a {\sinh \left( \tau  \right)}}} \right.
 \kern-\nulldelimiterspace} {\sinh \left( \tau  \right)}}}$ and center at $x={{x_B} + a\coth \left( \tau  \right)}$.
 To align $\partial {\mathcal{C}}_i$ with an iso-contour $\tau_i$ we thus require
\begin{subequations}
\begin{align}
{x_B} + a\coth \left( {{\tau _1}} \right) &=  - {d \mathord{\left/
{\vphantom {d 2}} \right.
\kern-\nulldelimiterspace} 2},\\
{a \mathord{\left/
{\vphantom {a {\sinh \left( {{\tau _1}} \right)}}} \right.
\kern-\nulldelimiterspace} {\sinh \left( {{\tau _1}} \right)}} &= r,\\
{x_B} + a\coth \left( {{\tau _2}} \right) &=   {d \mathord{\left/
{\vphantom {d 2}} \right.
\kern-\nulldelimiterspace} 2},\\
{a \mathord{\left/
{\vphantom {a {\sinh \left( {{\tau _2}} \right)}}} \right.
\kern-\nulldelimiterspace} {\sinh \left( {{\tau _2}} \right)}} &= 1,
\end{align}
\end{subequations}
which yields the solution \eqref{KinematicsEquationsRepere} for $x_B$, $a$ and  $\tau_i$.

The bipolar unit vectors ${{\bf{e}}_\sigma }$ and ${{\bf{e}}_\tau }$ shown in Fig. \ref{FigBipolarCoor} are expressed in the basis $\left( {{{\bf{e}}_x},{{\bf{e}}_y}} \right)$ as
\begin{subequations}
\begin{align}
{{\bf{e}}_\sigma } = \frac{1}{{\left| {{{\partial \;{\bf{g}}} \mathord{\left/
 {\vphantom {{\partial \;{\bf{g}}} {\partial \sigma }}} \right.
 \kern-\nulldelimiterspace} {\partial \sigma }}} \right|}}\frac{{\partial \;{\bf{g}}}}{{\partial \sigma }} = \frac{1}{{\cosh \left( \tau  \right) - \cos \left( \sigma  \right)}}{\left( {\begin{array}{*{20}{c}}
{ - \sin \left( \sigma  \right)\sinh \left( \tau  \right)}\\
{\cos \left( \sigma  \right)\cosh \left( \tau  \right) - 1}
\end{array}} \right)},\\
{{\bf{e}}_\tau } = \frac{1}{{\left| {{{\partial \;{\bf{g}}} \mathord{\left/
 {\vphantom {{\partial \;{\bf{g}}} {\partial \sigma }}} \right.
 \kern-\nulldelimiterspace} {\partial \sigma }}} \right|}}\frac{{\partial \;{\bf{g}}}}{{\partial \tau }} = \frac{1}{{\cosh \left( \tau  \right) - \cos \left( \sigma  \right)}}{\left( {\begin{array}{*{20}{c}}
{1 - \cos \left( \sigma  \right)\cosh \left( \tau  \right)}\\
{ - \sin \left( \sigma  \right)\sinh \left( \tau  \right)}
\end{array}} \right)}.
\end{align}
\end{subequations}

\section{Determination of the fluid potential}\label{AppendixPotential}
\subsubsection{Determination of ${\varphi _x}$}\label{Subsection_Phix}
The functions ${\varphi _x}$ and ${\varphi _y}$ are determined with a separation of variables method with a $2\pi$ periodic function of $\sigma$, \cite{Alassar2009}.

For ${\dot{\mathop u} _{iy}} = 0$, the fluid eqs. \eqref{LaplaceEquationBipolar} and \eqref{BoundaryInf} are automatically satisfied by the potential function
\begin{align}
{\varphi _x} &={\dot{\mathop u}_{1x}}\sum\limits_{n = 1}^\infty  {{\varphi _{1xn}}\left[\cos \left( {n\sigma } \right)\cosh \left( {n\left( {\tau  - {\tau _2}} \right)} \right)-   \cosh \left( {n{\tau _2}} \right)\right]}\nonumber\\
&+ {\dot{\mathop u}_{2x}}\sum\limits_{n = 1}^\infty  {{\varphi _{2xn}}\left[\cos \left( {n\sigma } \right)\cosh \left( {n\left( {\tau  - {\tau _1}} \right)} \right)-   \cosh \left( {n{\tau _1}} \right)\right]}.
\end{align}
The first terms in the brackets are harmonic, $2\pi$ periodic in $\sigma$ and yield a zero fluid velocity at one of the cylinder boundary.
The second constant terms are added to ensure that ${\varphi _x}$ vanishes at infinity, i.e. as $\left( {\sigma ,\tau } \right) \to \left( {0,0} \right)$.

The boundary condition \eqref{BoundaryFixed} yields two equations for  ${\varphi _{ixn}}$
\begin{subequations}
\begin{align}
\sum\limits_{n = 1}^\infty  {{\varphi _{1xn}}n\cos \left( {n\sigma } \right)\sinh \left( {n\left( {{\tau _1} - {\tau _2}} \right)} \right)}  = {g_{1x}},\\
\sum\limits_{n = 1}^\infty  {{\varphi _{2xn}}n\cos \left( {n\sigma } \right)\sinh \left( {n\left( {{\tau _2} - {\tau _1}} \right)} \right)}  = {g_{2x}},
\end{align}
\end{subequations}
whose solutions are obtained by taking the scalar product
\begin{equation}
\left\langle {f,h} \right\rangle  = \frac{1}{\pi }\int\limits_0^{2\pi } {f\left( \sigma  \right)} h\left( \sigma  \right)d\sigma ,
\end{equation}
of both sides by  $\cos \left( {n\sigma } \right)$, leading to
\begin{subequations}\label{Appendix_eq_phiIN}
\begin{align}
{\varphi _{1xn}} &= \frac{{\left\langle {\cos \left( {n\sigma } \right),{g_{1x}}\left( \sigma  \right)} \right\rangle }}{{n\sinh \left( {n\left( {{\tau _1} - {\tau _2}} \right)} \right)}} = \frac{{ - 2na{e^{ n{\tau _1}}}}}{{n\sinh \left( {n\left( {{\tau _1} - {\tau _2}} \right)} \right)}} ={\phi _{1n}},\\
{\varphi _{2xn}} &= \frac{{\left\langle {\cos \left( {n\sigma } \right),{g_{2x}}\left( \sigma  \right)} \right\rangle }}{{n\sinh \left( {n\left( {{\tau _2} - {\tau _1}} \right)} \right)}} = \frac{{ - 2na{e^{ - n{\tau _2}}}}}{{n\sinh \left( {n\left( {{\tau _2} - {\tau _1}} \right)} \right)}} ={\phi _{2n}}.
\end{align}
\end{subequations}

\subsubsection{Determination of ${\varphi _y}$}\label{Subsection_Phiy}
For ${\dot{\mathop u} _{ix}} = 0$, the fluid eqs. \eqref{LaplaceEquationBipolar} and \eqref{BoundaryInf} are automatically satisfied by the potential function
\begin{align}
{\varphi _y} &= {\dot{\mathop u}  _{1y}}\sum\limits_{n = 1}^\infty  {{\varphi _{1yn}}\sin \left( {n\sigma } \right)\cosh \left( {n\left( {\tau  - {\tau _2}} \right)} \right)}\nonumber\\
&+{\dot{\mathop u}  _{2y}}\sum\limits_{n = 1}^\infty  {{\varphi _{2yn}}\sin \left( {n\sigma } \right)\cosh \left( {n\left( {\tau  - {\tau _1}} \right)} \right)} .
\end{align}

The boundary condition \eqref{BoundaryFixed} yields two equations for  ${\varphi _{iyn}}$
\begin{subequations}
\begin{align}
\sum\limits_{n = 1}^\infty  {{\varphi _{1yn}}n\sin \left( {n\sigma } \right)\sinh \left( {n\left( {{\tau _1} - {\tau _2}} \right)} \right)}  &= {g_{1y}},\\
\sum\limits_{n = 1}^\infty  {{\varphi _{2yn}}n\sin \left( {n\sigma } \right)\sinh \left( {n\left( {{\tau _2} - {\tau _1}} \right)} \right)}  &= {g_{2y}},
\end{align}
\end{subequations}
whose solution are obtained by taking the dot product of both sides by $\sin \left( {n\sigma } \right)$
\begin{subequations}
\begin{align}
{\varphi _{1yn}} =\frac{{\left\langle {\sin \left( {n\sigma } \right),{g_{1y}}\left( \sigma  \right)} \right\rangle }}{{n\sinh \left( {n\left( {{\tau _1} - {\tau _2}} \right)} \right)}} = \frac{{ - 2na{e^{ n{\tau _1}}}}}{{n\sinh \left( {n\left( {{\tau _1} - {\tau _2}} \right)} \right)}} = {\phi _{1n}},\\
{\varphi _{2yn}} =\frac{{\left\langle {\sin \left( {n\sigma } \right),{g_{2y}}\left( \sigma  \right)} \right\rangle }}{{n\sinh \left( {n\left( {{\tau _2} - {\tau _1}} \right)} \right)}} = \frac{{ - 2na{e^{ - n{\tau _2}}}}}{{n\sinh \left( {n\left( {{\tau _2} - {\tau _1}} \right)} \right)}} = {\phi _{2n}}.
\end{align}
\end{subequations}


\bibliographystyle{elsarticle-num}
\bibliography{biblio}

\begin{thebibliography}{10}
\expandafter\ifx\csname url\endcsname\relax
  \def\url#1{\texttt{#1}}\fi
\expandafter\ifx\csname urlprefix\endcsname\relax\def\urlprefix{URL }\fi
\expandafter\ifx\csname href\endcsname\relax
  \def\href#1#2{#2} \def\path#1{#1}\fi

\bibitem{Furber1979}
S.~B. Furber, J.~E. Ffowcswilliams, Is the weis-fogh principle exploitable in
  turbomachinery?, Journal of Fluid Mechanics 94 (1979) 519--540.

\bibitem{Nair2007}
S.~Nair, E.~Kanso, Hydrodynamically coupled rigid bodies, Journal of Fluid
  Mechanics 592 (2007) 393--411.

\bibitem{Chen1975b}
S.~S. Chen, Vibration of nuclear fuel bundles, Nucl. Eng. Des. 35 (1975)
  399--422.

\bibitem{Chen1977}
S.~S. Chen, Dynamics of heat exchanger tube banks, J. Fluid. Eng. 99 (1977)
  462--469.

\bibitem{LeCunff2002}
C.~Le~Cunff, F.~Biolley, E.~Fontaine, S.~Etienne, M.~L. Facchinetti,
  Vortex-induced vibrations of risers: theoretical, numerical and experimental
  investigation, Oil {\&} Gas Science and Technology 57 (2002) 59--69.

\bibitem{DeLangre2008}
E.~De~Langre, Effects of wind on plants, Annual Review of Fluid Mechanics 40
  (2008) 141--168.

\bibitem{Doare2011}
O.~Doare, S.~Michelin, Piezoelectric coupling in energy-harvesting fluttering
  flexible plates: linear stability analysis and conversion efficiency, J.
  Fluids Struct. 27 (2011) 1357--1375.

\bibitem{Singh2012}
K.~Singh, S.~Michelin, E.~De~Langre, Energy harvesting from axial fluid-elastic
  instabilities of a cylinder, Journal of Fluids and Structures 30 (2012) 159
  -- 172.

\bibitem{Michelin2013}
S.~Michelin, O.~Doare, Energy harvesting efficiency of piezoelectric flags in
  axial flows, J. Fluid Mech. 714 (2013) 489--504.

\bibitem{Virot2016}
E.~Virot, X.~Amandolese, P.~Hemon, Coupling between a flag and a spring-mass
  oscillator, Journal of Fluids and Structures 65 (2016) 447--454.

\bibitem{Eloy2008}
C.~Eloy, R.~Lagrange, C.~Souilliez, L.~Schouveiler, Aeroelastic instability of
  cantilevered flexible plates in uniform flow, Journal of Fluid Mechanics 611
  (2008) 97--106.

\bibitem{Paidoussis2014}
M.~P. Paidoussis, S.~J. Price, E.~De~Langre, Fluid-structure interactions:
  cross-flow-induced instabilities, Cambridge University Press, 2014.

\bibitem{Chen1987}
S.~S. Chen, Flow-Induced Vibration of Circular Cylindrical Structures,
  Hemisphere Publishing, 1987.

\bibitem{Hicks1879}
W.~M. Hicks, On the motion of two cylinders in a fluid, The Quarterly Journal
  of Pure and Applied Mathematics 16 (1879) 113--140, 193--219.

\bibitem{Lamb1945}
H.~Lamb, Hydrodynamics, Dover, New York, 6nd ed., 1945.

\bibitem{Greenhill1882}
A.~G. Greenhill, Functional images in cartesians, The Quaterly Journal of Pure
  and Applied Mathematics 18 (182) 356--362.

\bibitem{Basset1888}
A.~B. Basset, A Treatise on Hydrodynamics, Deighton, Bell and co., 1888.

\bibitem{Carpenter1958}
L.~H. Carpenter, On the motion of two cylinders in an ideal fluid, J. Res. Nat.
  Bur. Stand. 61 (1958) 83--87.

\bibitem{Birkhoff1960}
G.~Birkhoff, Hydrodynamics, Princeton University Press, Princeton, New Jersey,
  2nd ed., 1960.

\bibitem{Gibert1980}
R.~J. Gibert, M.~Sagner, Vibration of structures in a static fluid medium, La
  Houille Blanche 1/2 (1980) 204--262.

\bibitem{Landweber1991}
L.~Landweber, A.~Shahshahan, Added masses and forces on two bodies approaching
  central impact in an inviscid fluid, in: Technical Rep. 346, Iowa Institute
  of Hydraulic Research, 1991.

\bibitem{Wang2004}
Q.~X. Wang, Interaction of two circular cylinders in inviscid fluid, Phys.
  Fluids 16 (2004) 4412.

\bibitem{Burton2004}
D.~A. Burton, J.~Gratus, R.~W. Tucker, Hydrodynamic forces on two moving discs,
  Theoret. Appl. Mech. 31 (2004) 153--188.

\bibitem{Tchieu2010}
A.~A. Tchieu, D.~Crowdy, A.~Leonard, Fluid-structure interaction of two bodies
  in an inviscid fluid, Phys. Fluids 22 (2010) 107101.

\bibitem{Scolan2008}
Y.~M. Scolan, S.~Etienne, On the use of conformal mapping for the computation
  of hydrodynamic forces acting on bodies of arbitrary shape in viscous flow.
  part 2: multi-body configuration, Journal of Engineering Mathematics 61
  (2008) 17--34.

\bibitem{Crowdy2006}
D.~G. Crowdy, Analytical solutions for uniform potential flow past multiple
  cylinders, Eur. J. Mech. B/Fluids 25 (2006) 459--470.

\bibitem{Crowdy2010}
D.~G. Crowdy, A new calculus for two-dimensional vortex dynamics, Theor.
  Comput. Fluid Dyn. 24 (2010) 9.

\bibitem{Morison1950}
J.~R. Morison, M.~P. O~Brien, J.~W. Johnson, S.~A. Schaaf, The force exerted by
  surface waves on piles, Petroleum Transactions, American Institute of Mining
  Engineers 189 (1950) 149--154.

\bibitem{Mazur1970}
V.~Y. Mazur, Motion of two circular cylinders in an ideal fluid, Izvestiya
  Akademii Nauk SSSR, Mekhanika Zhidkosti i Gaza 6 (1970) 80--84.

\bibitem{Herman1887}
R.~A. Herman, On the motion of two spheres in fluid and allied problem, The
  Quarterly Journal of Pure and Applied Mathematics 22 (1887) 204--262.

\bibitem{Paidoussis1984}
M.~P. Pa\"idoussis, D.~Mavriplis, S.~J. Price, A potential-flow theory for the
  dynamics of cylinder arrays in cross-flow, Journal of Fluid Mechanics 146
  (1984) 227--252.

\bibitem{Chen1978}
S.~S. Chen, Crossflow-induced vibrations of heat exchanger tube banks, Nuclear
  Engineering and Design 47 (1978) 67--86.

\bibitem{Suss1977}
S.~Suss, Stability of a cluster of flexible cylinders in bounded axial flow,
  Ph.D. thesis, McGill (1977).

\bibitem{Dalton1971}
C.~Dalton, R.~A. Helfinstine, Potential flow past a group of circular
  cylinders, Trans. ASME Journal of Basic Engineering 93 (1971) 636--642.

\bibitem{Bampalas2008}
N.~Bampalas, J.~M.~R. Graham, Flow-induced forces arising during the impact of
  two circular cylinders, J. Fluid Mech. 616 (2008) 205--234.

\bibitem{Alassar2009}
R.~S. Alassar, M.~A. El-Gebeily, Inviscid flow past two cylinders, ASME Trans.
  J. Fluid Eng. 131 (2009) 054501.

\end{thebibliography}




\end{document}